\documentclass[journal=jpclcd,manuscript=letter]{achemso}

\usepackage[version=3]{mhchem} 
\usepackage{amssymb}
\usepackage{amsmath}
\usepackage{gensymb}
\usepackage{bm}
\usepackage{soul,color}

\author{Xinqiang Ding}
\author{Bin Zhang}
\email{binz@mit.edu}
\affiliation[Massachusetts Institute of Technology]
{ Department of Chemistry, Massachusetts Institute of Technology, Cambridge, Massachusetts 02139, USA}

\title{Computing Absolute Free Energy with Deep Generative Models}


\begin{document}
\begin{tocentry}





\includegraphics[width=0.9\textwidth]{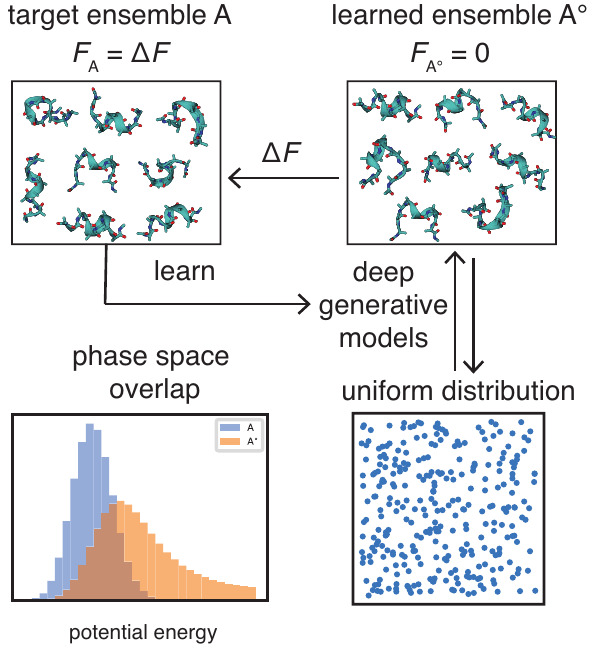}
\end{tocentry}

\begin{abstract}
     Fast and accurate evaluation of free energy has broad applications from drug design to material engineering. Computing the absolute free energy is of particular interest since it allows the assessment of the relative stability between states without intermediates. In this letter, we introduce a general framework for calculating the absolute free energy of a state. A key step of the calculation is the definition of a reference state with tractable deep generative models using locally sampled configurations. The absolute free energy of this reference state is zero by design. The free energy for the state of interest can then be determined as the difference from the reference. We applied this approach to both discrete and continuous systems and demonstrated its effectiveness. It was found that the Bennett acceptance ratio method provides more accurate and efficient free energy estimations than approximate expressions based on work. We anticipate the method presented here to be a valuable strategy for computing free energy differences.
\end{abstract}

Free energy is of central importance in both statistical physics and computational chemistry. It has important applications in rational drug design \cite{RamiReddy2001FreeDesign} and material property prediction \cite{Auer2001PredictionColloids}. Therefore, methodology development for efficient free energy calculations has attracted great research interest \cite{Torrie1977NonphysicalSampling,Kumar1992THEMethodb,Jorgensen1985MonteHydration,Shirts2008StatisticallyStatesb,Schneider2017StochasticSurfaces,Pohorille2010GoodCalculations,Klimovich2015GuidelinesCalculationsb,Kollman1993FreePhenomena,Hahn2009UsingEstimates,Jarzynski2002TargetedPerturbation,Wirnsberger2020TargetedMappings}. Many existing algorithms have focused on estimating free energy differences between states and originate from the free energy perturbation (FEP) identity \cite{Zwanzig1954High-TemperatureGases}
\begin{eqnarray}
    \mathbb{E}_{A}[e^{-\beta \Delta U}] = e^{-\beta \Delta F}.
    \label{equ:FEP}
\end{eqnarray}
Here, $\Delta F = F_B - F_A$ is the free energy difference between two equilibrium states $A$ and $B$ at temperature $T$ and $\beta = 1/k_B T$. $U_A(\bm{x})$ and $U_B(\bm{x})$ are the potential energies for a configuration $\bm{x}$ in states $A$ and $B$, respectively, and $\Delta U(\bm{x}) = U_B(\bm{x}) - U_A(\bm{x})$. $\mathbb{E}_A$ represents the expectation with respect to the Boltzmann distribution of $\bm{x}$ in state $A$,
\begin{eqnarray}
  p_A(\bm{x}) = \frac{e^{-\beta U_A(\bm{x})}}{Z_A},
\end{eqnarray}
where the normalization constant $Z_A=\int e^{-\beta U_A(\bm{x})} d\bm{x}$. Computing $\Delta F$ with the FEP identity (Eq. \ref{equ:FEP}) only uses samples from state $A$. It is more efficient to use samples from both states to compute $\Delta F$ by solving the Bennett acceptance ratio (BAR) equation  \cite{Bennett1976EfficientData}
\begin{eqnarray}
\label{equ:BAR}
  && \sum_{k=1}^{N_A}{f(\beta[\Delta U(\bm{x}^A_k) - M - \Delta F])}  \nonumber \\
  = && \sum_{k=1}^{N_B}{f(-\beta[\Delta U(\bm{x}^B_k) - M - \Delta F]))},
\end{eqnarray}
where $f(t) = 1/(1 + e^t)$ and $M = \ln (N_B/N_A)$. Here, $\{\bm{x}^A_k, k = 1, ..., N_A\}$ and $\{\bm{x}^B_k, k = 1, ..., N_B\}$ are samples from the two states. Both the FEP and the BAR method converge poorly when the overlap in the configuration space between state $A$ and $B$ is small. In that case, multiple intermediate states along a path with incremental changes in the configuration space can be introduced to bridge the two states \cite{Torrie1977NonphysicalSampling}. However, sampling from multiple intermediate states greatly increases the computational cost. It is, therefore, useful to develop techniques that can alleviate the convergence issue without the use of intermediate states \cite{Jarzynski2002TargetedPerturbation, Wirnsberger2020TargetedMappings}.

The requirement on a significant overlap between the two states' configuration space can be circumvented if we compute their free energy difference from the absolute free energy as $\Delta F = F_B - F_A $. The absolute free energy of a state $A/B$ can be obtained from its difference from a reference state $A^{\circ}/B^{\circ}$ as $F_{A/B} = F_{A^{\circ}/B^{\circ}} - \Delta F_{A/B \rightarrow A^{\circ}/B^{\circ}}$. For this strategy to be efficient, however, the reference states must bear significant overlap in configuration space with the states of interest. Their absolute free energy should be available with minimal computational effort. For most systems, designing reference states that satisfy these constraints can be challenging and requires expertise and physical intuition \cite{Hoover1971ThermodynamicPotentials,Frenkel1984NewSpheres,Hoover1967UsePhase,Amon2000DevelopmentStates,Ytreberg2006SimpleBiomolecules,schilling2009computing,berryman2013}. In this letter, we demonstrate that reference states can be constructed with tractable generative models for efficient computation of the absolute free energy \cite{Uria2016NeuralEstimation,Dinh2016DensityNvp}.

The workflow for calculating the absolute free energy is as follows. State $A$ is used as an example for the discussion, but the same procedure applies to state $B$. We first draw samples, $\{\bm{x}^A_k, k = 1, ..., N_A\}$, from the Boltzmann distribution $p_A(\bm{x})$. We then learn a tractable generative model, $q_\theta(\bm{x})$, that maximizes the likelihood of observing these samples by fine-tuning the set of parameters $\theta$. Here tractable generative models refer to probabilistic models that have the following two properties: (i) the normalized probability (or probability density), $q_\theta(\bm{x})$, can be directly evaluated for a given configuration $\bm{x}$ without the need of sampling or integration; (ii) independent configurations can be efficiently sampled from the probability distribution. 
The generative model defines a new equilibrium state $A^{\circ}$, which serves as an excellent reference to state $A$.  Because it is parameterized from samples of state $A$, most probable configurations from $A^{\circ}$ should resemble those from $A$ by design, and the overlap between the two states is guaranteed as long as the generative model has enough flexibility for modeling $p_A(\bm{x})$. In addition, since $q_\theta(\bm{x})$ is normalized, if we define the potential energy of state $A^{\circ}$ as  $ U_{A^{\circ}}(\bm{x}) = - (1/\beta) \ln q_\theta(\bm{x})$,
the partition function $Z_{A^{\circ}}$ of state $A^{\circ}$ is equal to 1, i.e.,
$Z_{A^{\circ}} = \int q_\theta(\bm{x}) \text{d}\bm{x} = 1$.
The absolute free energy of the reference state $A^{\circ}$ is $F_{A^{\circ}} = -(1/\beta) \ln Z_{A^{\circ}} = 0$. 
(Strictly speaking, the free energy should be defined as $F_{A^{\circ}} = -(1/\beta) \ln (Z_{A^{\circ}} / \int 1 \text{d}\bm{x})$ to normalize the unit in the partition function. This technical detail does not affect any of the conclusions on free energy differences and is not considered for simplicity.) With the reference state defined, the absolute free energy for state $A$ can be determined by solving a similar BAR equation as Eq. \ref{equ:BAR}. Our use of tractable generative models ensures that sample configurations can be easily produced for the reference state to be combined with those from state $A$ for solving the BAR equation.

We note that a closely related algorithm for computing the absolute free energy has been introduced in variational methods \cite{Opper2001AdvancedPractice,Wu2019SolvingNetworksb,Li2018NeuralGroup}. In these prior studies, $q_\theta(\bm{x})$ was optimized by minimizing the Kullback-Leibler (KL) divergence \cite{Kullback1951OnSufficiency} from $q_\theta(\bm{x})$ to $p_A(\bm{x})$
\begin{eqnarray}
\label{equ:dkl_qp}
D_{\text{KL}}(q_\theta||p_A) && = \int{q_\theta(\bm{x}) \ln \frac{q_\theta(\bm{x})}{p_A(\bm{x})}} \text{d}\bm{x} \nonumber \\
&&= \beta (\langle W_{A^{\circ} \rightarrow A} \rangle - F_A),
\end{eqnarray}
where $\langle W_{A^{\circ} \rightarrow A} \rangle = \mathbb{E}_{A^{\circ}}[U_A(\bm{x}) - U_{A^{\circ}}(\bm{x})]$.
Because $D_{\text{KL}}(q_\theta||p_A)$ is non-negative, $\langle W_{A^{\circ} \rightarrow A} \rangle$ is an upper bound of $F_A$. As $D_{\text{KL}}(q_\theta||p_A)$ decreases along the optimization, $\langle W_{A^{\circ} \rightarrow A}\rangle$ is assumed to approach closer to the true free energy and was used for its estimation.

Our methodology is different from the variational methods in two aspects.
Firstly, instead of $D_{\text{KL}}(q_\theta||p_A)$, we used
\begin{eqnarray}
\label{equ:dkl_pq}
D_{\text{KL}}(p_A||q_\theta) && = \int{p_A(\bm{x}) \ln \frac{p_A(\bm{x})}{q_\theta(\bm{x})}} \text{d}\bm{x} \nonumber \\
&& = \beta (\langle W_{A \rightarrow A^{\circ}}\rangle + F_A)
\end{eqnarray}
as the objective function for learning $q_\theta(\bm{x})$. $\langle W_{A \rightarrow A^{\circ}} \rangle = \mathbb{E}_{A}[U_{A^{\circ}}(\bm{x}) - U_{A}(\bm{x})]$. We note that minimizing the KL divergence from $p_A(\bm{x})$ to $q_\theta(\bm{x})$ is equivalent to learning the generative model by maximizing its likelihood on the training data. Moreover, because $D_{\text{KL}}(p_A||q_\theta)$ is also non-negative, $\langle -W_{A \rightarrow A^{\circ}} \rangle$ is a lower bound of $F_A$. Therefore, minimizing $D_{\text{KL}}(p_A||q_\theta)$ is equivalent to maximizing the lower bound $\langle -W_{A \rightarrow A^{\circ}} \rangle$. At the face value, it may seem that $D_{\text{KL}}(q_\theta||p_A)$ is a better objective function than $D_{\text{KL}}(p_A||q_\theta)$ for model training since its optimization only requires samples from $q_\theta(\bm{x})$. As aforementioned, sampling from $q_\theta(\bm{x})$ can be made computationally efficient by the use of tractable generative models. On the other hand, training by $D_{\text{KL}}(p_A||q_\theta)$ requires samples from $p_A(\bm{x})$, the collection of which often requires costly long timescale simulations with Monte Carlo or molecular dynamics (MD) techniques. The caveat is that optimization with $D_{\text{KL}}(q_\theta||p_A)$ is more susceptible to traps from local minima due to its more complex dependence on $q_\theta$. \cite{cover2006}
When $p_A(\bm{x})$ is a high-dimensional distribution and the system exhibits multistability, optimizing $D_{\text{KL}}(q_\theta||p_A)$ often leads to solutions that cover only one of the metastable states \cite{Noe2019BoltzmannLearning,wu2020stochastic}.
No\'{e} and coworkers have recognized the above challenge \cite{Noe2019BoltzmannLearning}, and they introduced the Boltzmann generator that uses a combination of both $D_{\text{KL}}(q_\theta||p_A)$ and $D_{\text{KL}}(p_A||q_\theta)$ for model training.

Another significant difference between our methodology and the variational methods or the Boltzmann generator is the expression used to estimate $ F_A $.  In particular, $\langle W_{A^{\circ} \rightarrow A} \rangle$ is an upper bound of the free energy and only becomes exact when the probability distributions from generative models and the state of interest are identical. On the other hand, our use of the BAR equation (Eq.\@ \ref{equ:BAR}) relaxes this requirement, and $F_A$ can be accurately determined even if the model training is not perfect and there are significant differences between the two distributions. In all but trivial examples, we anticipate that the learning process does not converge exactly to the true distribution $p_A(\bm{x})$ due to its high dimensionality and complexity. The BAR estimation, which is asymptotically unbiased  \cite{Shirts2003EquilibriumMethods},  will be crucial to ensure the accuracy of free energy calculations.

\begin{figure}[t]
    \includegraphics[width=0.8\textwidth]{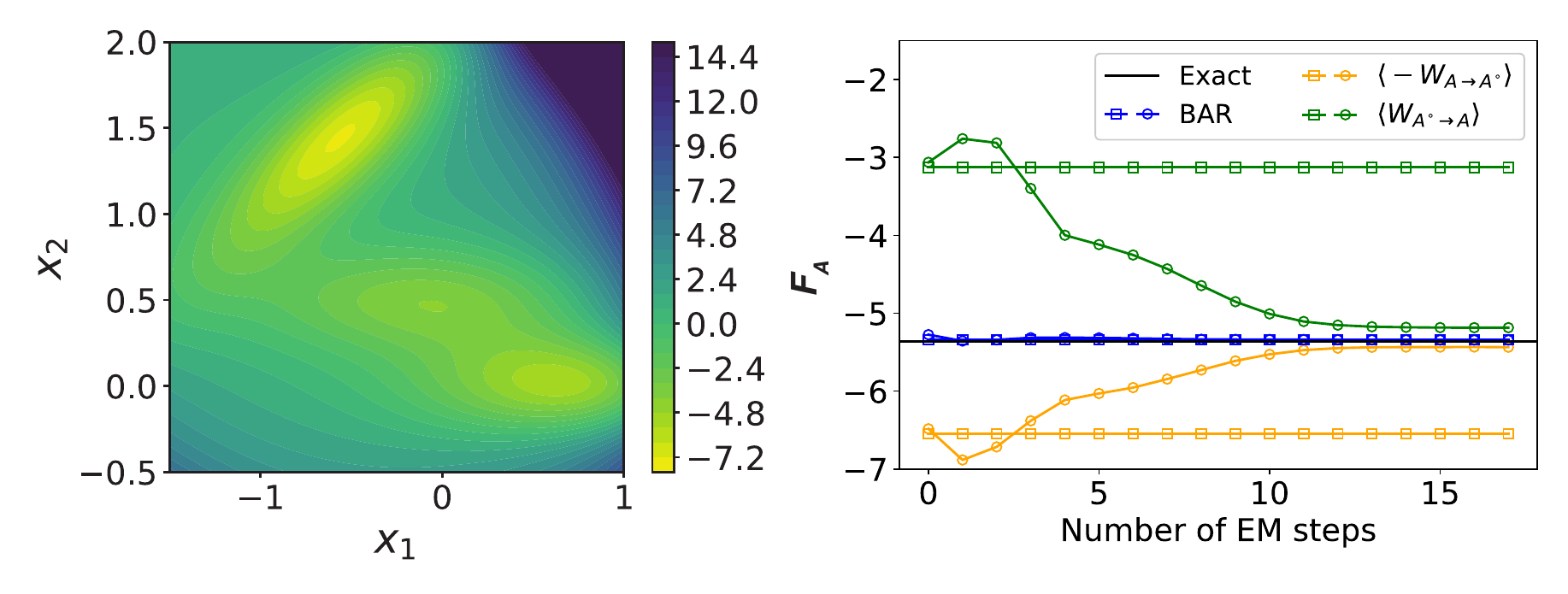}
    \caption{\label{fig:muller}
    (Left) Contour plot of the M\"{u}ller potential. Energy is shown in the units of $k_\mathrm{B}T$. (Right) Absolute free energy computed using various estimators with reference states defined as a Gaussian distribution (squares) or a mixture model of two Gaussian distributions (circles). 
    The $x$-axis corresponds to the number of steps for training the Gaussian mixture model with the expectation-maximization (EM) algorithm \cite{dempster1977maximum}.
}
\end{figure}

The advantage of the BAR estimation is evident when computing the absolute free energy of a two-dimensional system
with the M\"{u}ller potential \cite{muller1979location}. 
When the reference state $q_\theta(\bm{x})$ was parameterized with a Gaussian distribution, which fails to capture the multistability inherent to the system, the two bounds based on work deviate significantly from the absolute free energy (Fig.~\ref{fig:muller}). The free energy estimated using the BAR equation, on the other hand, is in excellent agreement with the exact value. The work-based bounds begin to approach the exact value when an optimized mixture model of two Gaussian distributions was used to parameterize the reference state. The BAR estimator again converges much faster than the bounds, highlighting its insensitivity to the quality of the reference state. 
More details about the model training and free energy computation for this simple test system are included in the Supplemental Material. 
We note that an independent study reported similar advantages when using BAR to compute relative free energy with deep generative models. \cite{Wirnsberger2020TargetedMappings}

\begin{figure}[t]
    \includegraphics[width=0.8\textwidth]{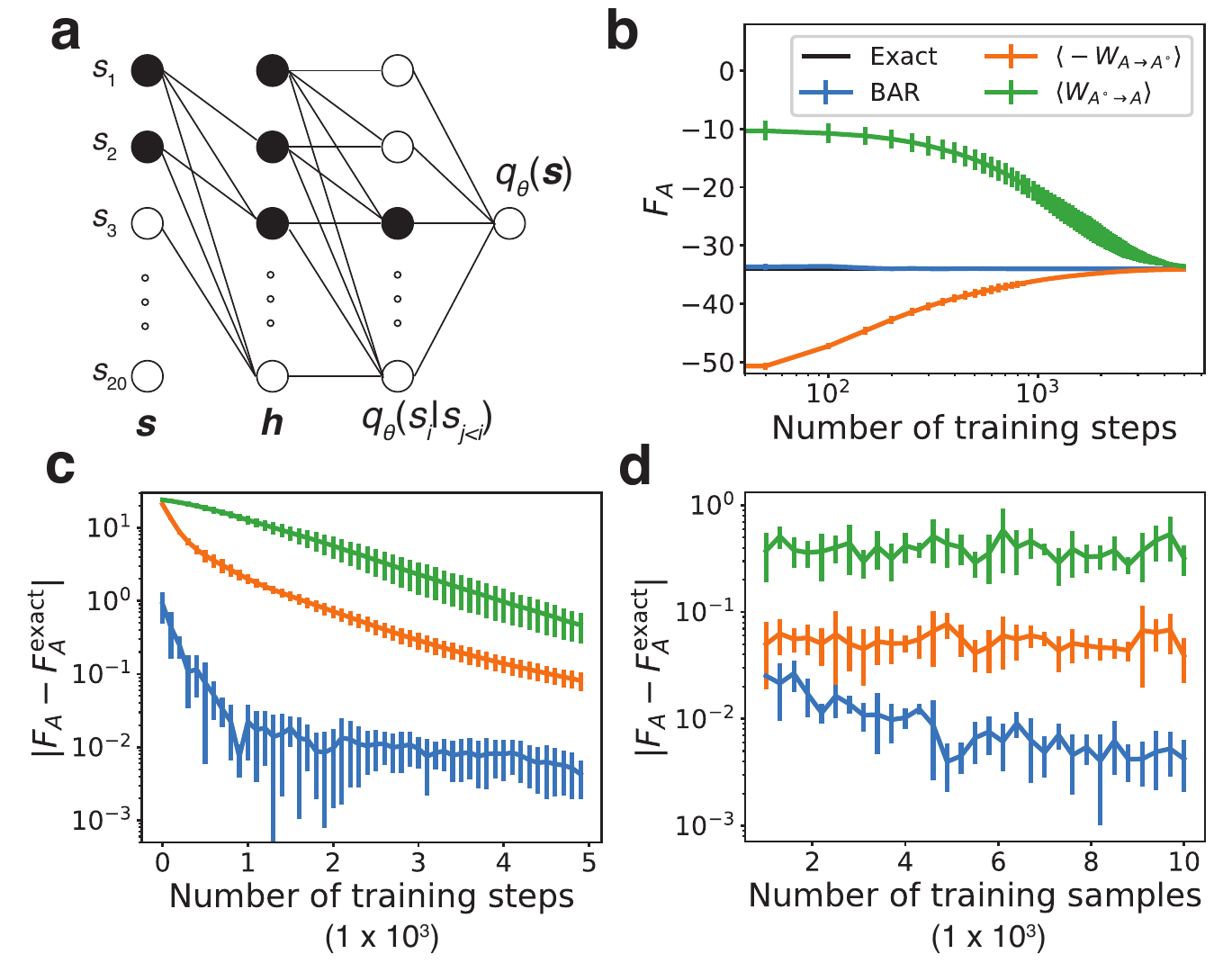}
    \caption{\label{fig:sk}
    Performance of different free energy estimators on the SK model. Energy is shown in the units of $k_\mathrm{B}T$. 
    (a) A schematic representation of the neural autoregressive model used to parameterize $q_\theta$. The black units illustrate the dependence of the conditional probability $q(s_3|s_1,s_2)$. 
    (b) The absolute free energy of the SK model calculated with three estimators as a function of training steps compared with the exact result obtained from a complete enumeration.
    (c, d) Errors of the estimated absolute free energy versus the number of training steps and the number of samples used for training $q_\theta$. The coloring scheme is identical to that in part b. 
    Error bars representing one standard deviation are estimated using five independent repeats.}
\end{figure}

Encouraged by the results from the above test system, we next computed the absolute free energy of a 20-spin classical Sherrington-Kerkpatrick (SK) model \cite{Sherrington1975SolvableSpin-glass}, the value of which can be determined from complete enumeration as well. The discrete configurations of the SK model will be represented using $\bm{s}$ instead of $\bm{x}$. Though we introduced the methodology with continuous variables, all the equations can be trivially extended to $\bm{s}$ by replacing the integrals with summations over the spin configurations. The potential energy of a configuration $\bm{s} = (s_1, s_2, ..., s_N)$ is defined as
\begin{eqnarray}
    U_A (\bm{s}) = \frac{1}{\sqrt{N}} \sum_{j>i} J_{ij} s_i s_j,
\end{eqnarray}
where $s_i \in \{-1, +1\}$ and $N = 20$. $J_{ij}$ were chosen randomly from the standard normal distribution. 5000 samples were drawn from the probability distribution $p(\bm{s}) = e^{-\beta U_A(\bm{s})}/Z_A$ with $\beta = 2.0$. These samples were used to train the reference state $A^{\circ}$ by minimizing $D_{\text{KL}}(p_A||q_\theta)$ (Eq.\@ \ref{equ:dkl_pq}). The reference probability $q_\theta(\bm{s})$ was defined with a neural autoregressive density estimator (NADE) \cite{Uria2016NeuralEstimation,Papamakarios2019NormalizingInferenceb,NIPS2016_6581,Papamakarios2017MaskedEstimation,Huang2018NeuralFlows} as a product of conditional distributions
\begin{eqnarray}
    \label{equ:autoregressive}
    q_\theta(\bm{s}) = \prod_{i=1}^{N}{q_\theta(s_i|s_1, ..., s_{i-1})}.
\end{eqnarray}
$q_\theta(s_i|s_1, ..., s_{i-1})$ were parameterized using a feed-forward neural network with one hidden layer of 20 hidden units.
The neural network's connections are specifically designed such that it maintains the autoregressive property, i.e., $q_\theta(s_i|s_1, ..., s_{i-1})$ only depends on $s_1, ..., s_{i}$ (Fig.\@ \ref{fig:sk}a). 
After training $q_\theta(\bm{s})$ for some numbers of steps \cite{Kingma2015Adam:Optimization,Paszke2019AutomaticPyTorch}, 5000 configurations were independently drawn from $q_\theta(\bm{s})$. These configurations, together with the training inputs sampled from $p_A(\bm{s})$, were used to determine the absolute free energy of the SK model. In Figs.\@ \ref{fig:sk}b and \ref{fig:sk}c, we again compare results from the three estimators with the exact value.

Similar to the results observed for the M\"{u}ller potential, at early stages of model parameterization with small training step numbers, the work-based estimations deviate significantly from the true value. This deviation is expected and is a direct result of the difference between the two probability distributions $q_\theta (\bm{s})$ and $p_A(\bm{s})$. 
However, as the training proceeds, the agreement between the distributions improves and $\langle -W_{A \rightarrow A^{\circ}} \rangle$ and $\langle W_{A^{\circ} \rightarrow A} \rangle$ gradually converge to the exact result after 5000 steps (Fig.\@ \ref{fig:sk}b) because the autoregressive model is flexible enough to match the target distribution.
On the other hand, the BAR estimator converges much faster to the exact value with a smaller error (Figs.\@ \ref{fig:sk}b and \ref{fig:sk}c). 
In addition, varying the number of samples used for training $q_\theta (\bm{s})$ has different effects on the accuracy of converged results for the three approaches (Fig.\@ \ref{fig:sk}d). For both $\langle -W_{A \rightarrow A^{\circ}} \rangle$ and $\langle W_{A^{\circ} \rightarrow A} \rangle$, increasing the number of training samples from $10^3$ to $10^4$ does not significantly change the accuracy of their results. In contrast, using more training samples significantly reduces the error of the BAR estimator. This is because solutions of the BAR equation are asymptotically unbiased for estimating $F_A$, whereas $\langle -W_{A \rightarrow A^{\circ}} \rangle $ and $\langle W_{A^{\circ} \rightarrow A} \rangle$ are not \cite{Shirts2003EquilibriumMethods}.

\begin{figure}[tb]
\includegraphics[width=0.8\textwidth]{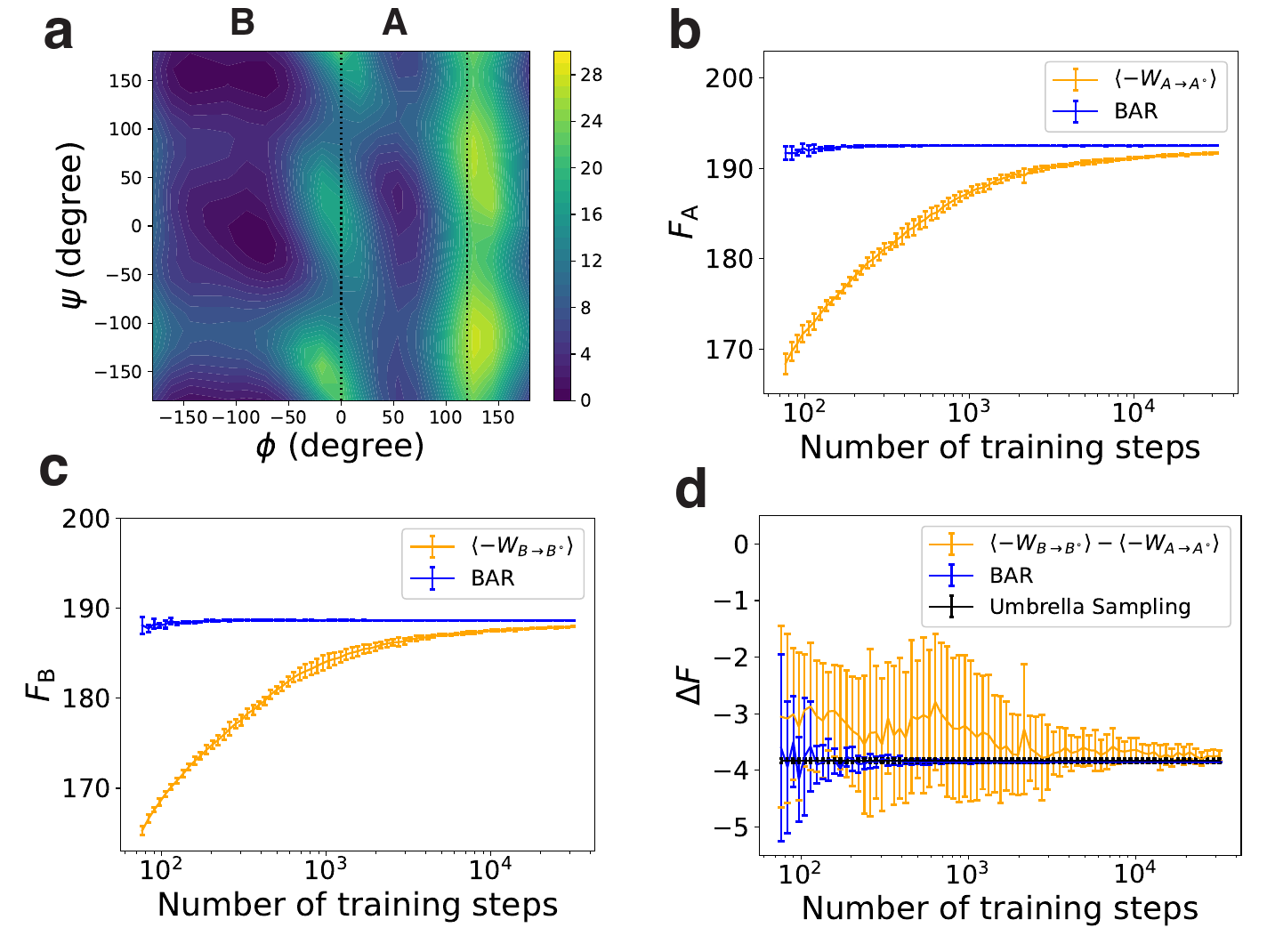} 
\caption{\label{fig:di-alanine} Performance of different free energy estimators on the di-alanine. Energy is shown in the units of $k_\mathrm{B}T$.
(a) Contour plot of the di-alanine free energy surface as a function of the two torsion angles $\phi$ and $\psi$.
(b, c) The absolute free energy of state $A$ and $B$ computed with different estimators.
(d) The free energy difference between state $A$ and $B$ computed with different estimators.
Error bars representing one standard deviation are estimated using five independent repeats.
}
\end{figure}

\begin{figure}[tb]
\includegraphics[width=0.8\textwidth]{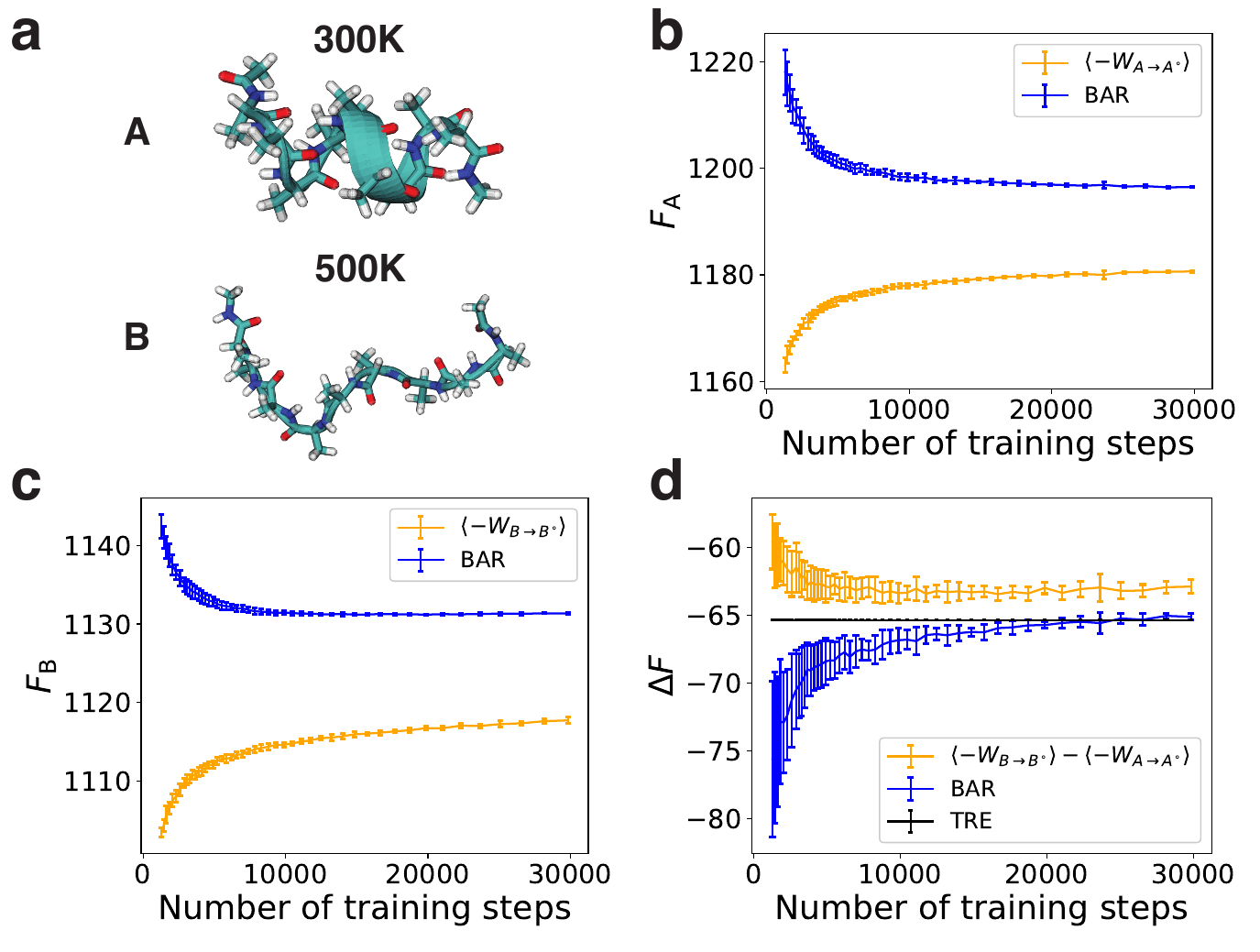} 
\caption{\label{fig:deca-alanine} Performance of different free energy estimators on the deca-alanine. Energy is shown in the units of $k_\mathrm{B}T$, with $T=300~K$.
(a) Representative conformations from state $A$ (the ensemble at $T=300K$) and state $B$  (the ensemble at $T=500K$).
(b, c) The absolute free energy of states A and B computed with different estimators.
(d) The free energy difference between states A and B computed with different estimators.
Error bars representing one standard deviation are estimated using five independent repeats.
}
\end{figure}

Finally, we applied the methodology to two molecular systems, the di-alanine and the deca-alanine in implicit solvent. These two systems present features commonly encountered in biomolecular simulations with continuous phase space over a rugged energy landscape. Their high dimensionality renders a complete enumeration of the configurational space to compute the absolute free energy for benchmarking impractical. Instead, we calculated the free energy difference between two metastable states using their absolute free energy to compare against the value determined from umbrella sampling and temperature replica exchange (TRE) simulations.
For di-alanine, the two metastable states were defined using the backbone dihedral angle $\phi$ (C-CA-N-C), with $0{\degree} < \phi \leq 120{\degree}$ for state $A$  and $\phi \leq 0{\degree}$ or $\phi > 120{\degree}$ for state $B$ (Fig.\@ \ref{fig:di-alanine}a).
For deca-alanine, states $A$ and $B$ were defined as the configurational ensembles at $T = 300K$ and $T = 500K$, respectively (Fig.\@ \ref{fig:deca-alanine}a).

To compute the absolute free energy, we learned the reference states using normalizing flow based generative models \cite{Rezende2015VariationalFlows,Papamakarios2019NormalizingInferenceb}. Specifically, $q_\theta(\bm{x})$ was parameterized with multiple bijective transformations, $T_1, ..., T_K$, to convert a random variable $\bm{u}$ to a peptide configuration, 
i.e.,
\begin{eqnarray}
    \bm{x} = T(\bm{u}) = T_K \circ \cdot\cdot\cdot \circ T_1(\bm{u}).
\end{eqnarray}
$\bm{u}$ shares the same dimension as $\bm{x}$ and is from a simple base distribution $p_u(\bm{u})$.
Based on the formula of variable change in probability density functions, we have
\begin{eqnarray}
    \label{equ:nf}
    \ln q_\theta(\bm{x}) = \ln p_u(\bm{u}) - \sum_{k = 1}^{K} {\ln |J_{T_k}(\bm{u}_{k-1})|},
\end{eqnarray}
where $\bm{u}_k = T_k \circ \cdot\cdot\cdot T_1(\bm{u}_0)$ and $\bm{u}_0 = \bm{u}$. $J_{T_k}$ is the Jacobian matrix of the transformation $T_k$, and $|\cdot|$ denotes the absolute value of the determinant.
For both molecules, we first transformed $\bm{u}$ into the internal coordinates $\bm{z}$ based on moleculear topology and then transformed $\bm{z}$ into the Cartesian coordinates $\bm{x}$ using the neural spline flows \cite{durkan2019neural,rezende2020normalizing} with coupling layers \cite{Dinh2016DensityNvp}.

The reference models were separately trained 
using configurations collected for each state from molecular dynamics simulations with the Amber ff99SB force field \cite{tian2019ff19sb} and the OBC implicit solvent model \cite{onufriev2004exploring}.
As shown in Figs.~S1-S4, they succeed in generating peptide conformations with reasonable geometry and energy (Figs.~S1-S4).
With the learned reference states, we computed the absolute free energy for states $A$ and $B$ using the three estimators. As shown in Figs.\@ \ref{fig:di-alanine} and \ref{fig:deca-alanine},
the BAR estimator converges much faster than the upper and lower bounds. The results calculated using the upper bound are not shown here because they are much larger than that of the lower bound and the BAR estimator (Figs.\@ S5 and S6). Unlike the results for the SK model, the two bounds no longer converge to the same value or the BAR estimator, and their difference can be as large as 6 $k_\mathrm{B}T$ for di-alanine (Fig.\@ S5) and 60 $k_\mathrm{B}T$ for deca-alanine (Fig.\@ S6). The large gaps between the two bounds suggest that the generative models are still quite different from the true distributions even after the learning has converged. We expect the numbers from the BAR estimator to be correct, because the BAR estimator does not require the generative models to precisely match the original distributions to reproduce the free energy, as shown in both the M\"{u}ller system and the SK model. Furthermore, the BAR estimations lie in between the two bounds in all four cases (Fig.\@ S5 and S6), as expected for the exact values. Therefore, for these two molecular systems, the two bounds cannot be used for reliable estimation of the absolute free energy.

We further evaluated the accuracy of the three estimators in computing the free energy differences between states $A$ and $B$. For comparison, we also determined the free energy difference using umbrella sampling \cite{Torrie1977NonphysicalSampling} for di-alanine and TRE simulations for deca-alanine. Results of estimated free energy differences are shown in Figs.\@ \ref{fig:di-alanine}d and \ref{fig:deca-alanine}d.
The BAR estimator converges much faster to the results from umbrella sampling or TRE simulations than the two bounds.
For di-alanine, the free energy difference estimated using BAR is $-3.86 \pm 0.01$ $k_\mathrm{B}T$, which agrees with the result from umbrella sampling ($-3.82 \pm 0.04$ $k_\mathrm{B}T$). 
To our surprise, the difference computed using the lower bound, $-3.74 \pm 0.10$ $k_\mathrm{B}T$, is close to the correct result as well.
Because the lower bound is biased,  we believe its good performance on the free energy difference is due to error cancellation.
For deca-alanine, the free energy difference from TRE is $-65.37 \pm 0.02$ $k_\mathrm{B}T$, which deviates from the result obtained from the lower bound ($-62.91 \pm 0.52$ $k_\mathrm{B}T$) but agrees well the BAR estimation ($-65.13 \pm 0.23$ $k_\mathrm{B}T$). 

In summary, we demonstrated that the framework based on deep generative models succeeds at computing the absolute free energy using sample configurations from the state of interest and is applicable for both discrete and continuous systems. It could greatly facilitate the evaluation of protein-ligand binding affinity and protein conformational stability while accounting for entropic contributions. Generalizing the methodology to 
compute the absolute free energy of biomolecular systems with explicit solvation \cite{Wirnsberger2020TargetedMappings,Noe2019BoltzmannLearning,Kohler2019EquivariantEnergies,Bender2019ExchangeableScans} would be an exciting direction for future studies.

\begin{acknowledgement}
This work was supported by the National Institutes of Health (Grant 1R35GM133580-01).
\end{acknowledgement}


\bibliography{references}
\end{document}